\documentstyle[12pt,epsf,epsfig]{article}
\newcount\timecount
\newcount\hours \newcount\minutes  \newcount\temp \newcount\pmhours
\hours = \time
\divide\hours by 60
\temp = \hours
\multiply\temp by 60
\minutes = \time
\advance\minutes by -\temp
\def\hour{\the\hours}
\def\minute{\ifnum\minutes<10 0\the\minutes
            \else\the\minutes\fi}
\def\clock{
\ifnum\hours=0 12:\minute\ AM
\else\ifnum\hours<12 \hour:\minute\ AM
      \else\ifnum\hours=12 12:\minute\ PM
            \else\ifnum\hours>12
                 \pmhours=\hours
                 \advance\pmhours by -12
                 \the\pmhours:\minute\ PM
                 \fi
            \fi
      \fi
\fi
}

\def\monthname{\relax\ifcase\month 0/\or January\or February\or
   March\or April\or May\or June\or July\or August\or September\or
   October\or November\or December\else\number\month/\fi}

\def\bold#1{\setbox0=\hbox{$#1$}%
     \kern-.025em\copy0\kern-\wd0
     \kern.05em\copy0\kern-\wd0
     \kern-.025em\raise.0433em\box0 }

\def\beq{\begin{equation}}
\def\eeq{\end{equation}}
\def\ba{\begin{eqnarray}}
\def\ea{\end{eqnarray}}

\def\ss{\scriptscriptstyle}
\def\ga{\mathrel{\raise.3ex\hbox{$>$\kern-.75em\lower1ex\hbox{$\sim$}}}}
\def\la{\mathrel{\raise.3ex\hbox{$<$\kern-.75em\lower1ex\hbox{$\sim$}}}}
\def\gev{{\rm \, Ge\kern-0.125em V}}
\def\tev{{\rm \, Te\kern-0.125em V}}
\def\gyr{{\rm \, G\kern-0.125em yr}}

\def\gappeq{\mathrel{\rlap {\raise.5ex\hbox{$>$}}
{\lower.5ex\hbox{$\sim$}}}}
\def\lappeq{\mathrel{\rlap{\raise.5ex\hbox{$<$}}
{\lower.5ex\hbox{$\sim$}}}}
\def\Toprel#1\over#2{\mathrel{\mathop{#2}\limits^{#1}}}

\def\stau{\widetilde \tau}
\def\stop{\widetilde t}

\def\m12{m_{1\!/2}}

\def\mz{m_{\ss Z}}

\begin{document}
\begin{titlepage}
\pagestyle{empty}
\baselineskip=21pt
\rightline{\tt hep-ph/0305212}
\rightline{CERN--TH/2003-107}
\rightline{UMN--TH--2201/03}
\rightline{FTPI--MINN--03/12}
\vskip 0.2in
\begin{center}
{\large {\bf Phenomenological Constraints on Patterns of Supersymmetry Breaking}}
\end{center}
\begin{center}
\vskip 0.2in
{\bf John~Ellis}$^1$, {\bf Keith~A.~Olive}$^{2}$, {\bf Yudi Santoso}$^{2}$
and {\bf Vassilis~C. Spanos}$^{2}$
\vskip 0.1in
{\it
$^1${TH Division, CERN, Geneva, Switzerland}\\
$^2${William I. Fine Theoretical Physics Institute, \\
University of Minnesota, Minneapolis, MN 55455, USA}}\\
\vskip 0.2in
{\bf Abstract}
\end{center}
\baselineskip=18pt \noindent

Specific models of supersymmetry breaking predict relations between the
trilinear and bilinear soft supersymmetry breaking parameters $A_0$ and
$B_0$ at the input scale. In such models, the value of $\tan \beta$ can be
calculated as a function of the scalar masses $m_0$ and the gaugino masses
$m_{1/2}$, which we assume to be universal. The experimental constraints
on sparticle and Higgs masses, $b \to s \gamma$ decay and the cold dark
matter density $\Omega_{CDM} h^2$ can then be used to constrain $\tan
\beta$ in such specific models of supersymmetry breaking. In the simplest
Polonyi model with $A_0 = (3 - \sqrt{3})m_0 = B_0 + m_0$, we find $11
\lappeq \tan \beta \lappeq 20$ ($\tan \beta \simeq 4.15$) for $\mu > 0$ ($\mu < 0$).
We also discuss other models with $A_0 = B_0 + m_0$, finding that only
the range $-1.9 \la A_0/m_0 \la 2.5$ is allowed for $\mu > 0$, and the
range $1.25 \la A_0/m_0 \la  4.8$ for $\mu < 0$. In these models,
we find no solutions in the rapid-annihilation `funnels' or in the `focus-point' region.
We also discuss the allowed
range of $\tan \beta$ in the no-scale model with $A_0 = B_0 = 0$. In all
these models, most of the allowed regions are in the $\chi - {\tilde
\tau_1}$ coannihilation `tail'. 
\vfill
\leftline{CERN--TH/2003-107}
\leftline{May 2003}
\end{titlepage}
\baselineskip=18pt

\section{Introduction}

One of the most important and least understood problems in the
construction of supersymmetric models is the mechanism of supersymmetry
breaking~\cite{BIM}. Direct exploration of this may be far beyond our 
experimental
reach for some considerable time, so we may have to rely on indirect
information provided by measurements of the different soft
supersymmetry-breaking parameters. Even here, so far we have no determinations,
only limits obtained from accelerator experiments, cosmology and
theoretical considerations. It is commonly assumed that the soft
supersymmetry-breaking scalar masses $m_0$ have universal values at some
GUT input scale, as do the gaugino masses $m_{1/2}$ and the trilinear soft
supersymmetry-breaking parameters $A_0$, which is referred to as the constrained
MSSM (CMSSM). One then frequently analyzes the impacts of the different 
phenomenological limits on the allowed values of $m_{1/2}$ and $m_0$ as 
functions of $\tan \beta$, the ratio of Higgs vacuum expectation values, 
assuming some default value of $A_0$ and determining the Higgs mixing 
parameter $\mu$ and the pseudoscalar Higgs mass $m_A$ by using the 
electroweak vacuum consistency 
conditions (see \cite{efgos} - \cite{hyperbolic} for recent studies of this type). 
The tree-level value of $m_A$ may be 
related to the bilinear soft supersymmetry-breaking parameter $B$, via 
$m_A^2 = - 2 B \mu / \sin 2 \beta$.

Specific models of supersymmetry breaking predict relations between these 
different soft supersymmetry-breaking parameters. For example, certain 
`no-scale' models \cite{noscale} may predict $m_0 = 0$ at the Planck scale, and we have 
analyzed the extent to which this assumption is compatible with the 
phenomenological constraints, taking account of the possible running of 
$m_0$ between the Planck scale and the GUT scale \cite{eno5}. Here we analyze a 
different question, namely the consistency of some proposed relations 
between $m_0$, $A_0$ and $B_0$ which take the characteristic form
\beq
A_0 \; = \; {\hat A} m_0, \quad \quad B_0 \; = \; {\hat B} m_0.
\label{hats}
\eeq
A generic minimal supergravity model \cite{sugr2} prediction is that
${\hat B} = {\hat A} -1$ \cite{mark}, and the simplest Polonyi 
model~\cite{pol} predicts that $\vert {\hat A} \vert = 3 - \sqrt{3}$~\cite{bfs}.

The first of the two relations (\ref{hats}) may be used to replace an {\it
ad hoc} assumption on the input value of $A_0$. The second imposes an
important consistency condition on the value of $m_A$, which was otherwise
treated as a dependent quantity that was not constrained {\it a priori}.  
For any given value of $m_{1/2}$ and $m_0$, this constraint is satisfied
for only one specific value of $\tan \beta$. Therefore, the results of
imposing the two constraints (\ref{hats}) may conveniently be displayed in 
a single $(m_{1/2}, m_0)$ plane across which $\tan \beta$ varies in a 
determined manner. The phenomenological constraints on $m_{1/2}$ and $m_0$ 
can then be used to provide both upper and lower limits on the allowed 
values of $\tan \beta$.

In this paper, we analyze these constraints on $\tan \beta$ as functions
of ${\hat A}$ in the generic scenario (\ref{hats}), including the Polonyi
case ${\hat A} = 3 - \sqrt{3}$ and other models with ${\hat A} = {\hat B}
+ 1$. In the Polonyi case, we find that $11 \lappeq \tan \beta \lappeq 20$
for $\mu > 0$, with only a small area in the $m_{1/2} - m_0$ plane with
$\tan \beta \simeq 4.15$ surviving for $\mu < 0$. In general, we find
consistent solutions for $-1.9 \la {\hat A} \la 2.25$ for $\mu > 0$ and
$1.25 \la {\hat A} \la 4.8$ for $\mu < 0$. We also explore the range of
$\tan \beta$ that is allowed in a no-scale scenario with $A_0 = B_0 = 0$
at the GUT scale. It should, however, be recalled that the no-scale
boundary conditions~\cite{noscale} were originally proposed to hold at the
supergravity scale, which might be significantly above the GUT scale. In
this case, renormalization-group running between these scales would
generate ${\hat A}$ and ${\hat B} \ne 0$ at the GUT scale.

\section{Models of Supersymmetry Breaking}

In this Section, we review briefly models that yield the characteristic
patterns of supersymmetry breaking whose phenomenology we study later in
the paper. We assume an $N = 1$ supergravity framework, interpreted as a
low-energy effective field theory. This may be characterized by a
K{\"a}hler function $K$ that describes the kinetic terms for the chiral
supermultiplets $\Phi \equiv (\zeta, \phi)$, where the $\zeta$ represent
hidden-sector fields and the $\phi^i$ observable-sector fields, a
holomorphic function $f(\Phi)$ that yields kinetic terms for the gauge
supermultiplets $A_a$ as well as gauge couplings, and a holomorphic
superpotential $W(\Phi)$. We assume the form of the gauge kinetic function
$f$ to be such that the gaugino masses $m_{1/2}$ are universal at the GUT
input scale, as are the gauge couplings.

So-called minimal supergravity theories have $K = \Sigma_i |\Phi^i|^2$,
whereas no-scale models have non-trivial K{\"a}hler functions such as $K =
- 3{\rm ln}(\zeta + \zeta^\dagger - \Sigma_j |\phi^j|^2)$. The scalar 
potential
(neglecting any gauge contributions) is in general~\cite{sugr2}
\beq
V(\phi,\phi^*) \; = \; e^K \left[ K^i (K^{-1})^j_i
K_j -3 \right]
\label{generalpot}
\eeq
where we are working in Planck units.
For minimal supergravity, we have $K^i = 
{\phi^i}^* + {W^i}/W$, $K_i =  \phi_i + W_i^*/W^*$, and
$({K^{-1}})^j_i = \delta ^j_i$, and the resulting scalar
potential is
\beq
V(\phi,\phi^*) \; = \; e^{ \phi_i {\phi^i}^*} \left[
|W^i + {\phi^i}^* W |^2 - 3|W|^2 \right].
\label{msgpot}
\eeq
In this minimal case, the soft supersymmetry-breaking
scalar masses $m_0$ are universal at the input GUT scale, with~\cite{BIM}
\beq
m_0^2 \; = \; m_{3/2}^2 + \Lambda,
\label{msugra}
\eeq
where $m_{3/2}$ is the gravitino mass and $\Lambda$ is the tree-level
cosmological constant. If we further assume that the
superpotential $W(\Phi)$ may be separated into pieces $F$ and $g$ that are
functions only of observable-sector fields $\phi^i$ and hidden-sector
fields $\zeta$, respectively, so that the superpotential
parameters of the observable-sector fields do not depend on the
hidden-sector fields, then the trilinear terms $A_0$ and bilinear terms
$B_0$ are also universal, and~\cite{BIM}
\beq
B_0 \; = \; A_0 - m_{3/2}.
\label{BA}
\eeq
Finally, if we further assume that $\Lambda = 0$, then $m_0 = m_{3/2}$ 
and~\cite{BIM}
\beq
{\hat B} \; = \; {\hat A} - 1,
\label{BAhat}
\eeq
which is one of the principal options we study below.

One of the primary motivations for the CMSSM, and for scalar mass
universality in particular, comes from the simplest model for local
supersymmetry breaking~\cite{pol}, which involves just one additional
chiral multiplet $\zeta$ in addition to the observable matter fields
$\phi_i$. We consider, therefore, a superpotential which is separable in
this so-called Polonyi field and the $\phi_i$, and of the simple form
\beq
g(\zeta) \; = \; \nu(\zeta + \beta)
\label{polonyi}
\eeq
with $\vert \beta \vert = 2 - \sqrt{3}$, ensuring that $\Lambda = 0$. The 
scalar 
potential in this model takes the form~\cite{bfs}
\ba
V & = & e^{(|\zeta|^2 + |\phi|^2)} \left[ |{\partial g \over \partial 
\zeta} + \zeta^* (g(\zeta) + F(\phi) )|^2 \right. \nonumber \\
& & + \left. |{\partial F \over \partial \phi} +
\phi^* (g(\zeta) + F(\phi) )|^2 - 3 | g(\zeta) + F(\phi) |^2 \right].
\label{cpot0}
\ea
We next expand the 
expression (\ref{cpot0}) and drop terms that are suppressed by 
inverse powers of the Planck scale, which can be done simply by dropping 
terms of mass dimension greater than four. In the positive case, after 
inserting the vev for
$\zeta$, $\langle \zeta \rangle = \sqrt{3} - 1$, we have~\cite{bfs}: 
\ba
V & = & e^{(4 - 2\sqrt{3})} \left[ |\nu + (\sqrt{3} -
1) (\nu + F(\phi)) |^2 \right. \nonumber \\
& & \left. +|{\partial F \over
\partial \phi} + \phi^* (\nu + F(\phi) )|^2 - 3 | \nu + F(\phi) |^2
\right] \nonumber \\
& = & e^{(4 - 2\sqrt{3})} |{\partial F \over \partial
\phi}|^2 \nonumber \\
& & + m_{3/2} e^{(2 - \sqrt{3})}(\phi {\partial F
\over \partial \phi} - \sqrt{3} F + h.c.) )  + m_{3/2}^2 \phi
\phi^* ,
\label{cpot}
\ea
which deserves some discussion.

First, up to an overall rescaling of the superpotential, $F \to
e^{\sqrt{3}-2} F$, the first term is the ordinary $F$-term part of the
scalar potential of global supersymmetry. The next term, which is
proportional to $m_{3/2}$, provides universal trilinear soft
supersymmetry-breaking terms $A = (3 -
\sqrt{3}) m_{3/2}$ and bilinear
soft supersymmetry-breaking terms $B = (2 - \sqrt{3}) m_{3/2}$, i.e., a
special case of the general relation (\ref{BA}) above between $B$ and
$A$. Finally, the last term represents a universal scalar mass of the type
advocated in the CMSSM, with $m_0^2 = m_{3/2}^2$, since the cosmological
constant $\Lambda$ vanishes in this model, by construction.

As we have seen above, the generation of such soft terms is a rather
generic property of low-energy supergravity models~\cite{mark} and many of
these conclusions persist when one generalizes the Polonyi potential. For
example, if we choose $g(\zeta)$ so that~\footnote{One could also consider 
models in which several fields $\zeta_i$ contribute to supersymmetry 
breaking.}
$\langle g \rangle = \nu$, $\langle \partial g / \partial \zeta \rangle =
a^* \nu$, and $\langle \zeta \rangle = b$, the condition that $\Lambda =
0$ at $\zeta = b$ implies $|a + b|^2 = 3$. Substituting these expectation
values in (\ref{cpot0}), we find~\cite{mark} that $A = b^* ( a+ b) \nu $
and once again $B = A - \nu$, but now with $A$ free.  The constant $\nu$ 
determines the gravitino mass, and hence $m_0$, through: 
$m_0 = m_{3/2} = e^{{1 \over 2} b b^*} \nu$.

Another broad option for supersymmetry breaking is that provided by
no-scale models \cite{noscale}, of which the simplest example is
\beq
K \; = \; - 3 {\rm ln} \left( \zeta + \zeta^\dagger - \Sigma_i | \phi^i
|^2 \right).
\label{noscale}
\eeq
No-scale models have the universal values
\beq
m^2_0 \; = \; 0, \; A_0 \; = \; 0, \; B_0 \; = \; 0
\label{noscalesusyx}
\eeq
at the input supergravity scale. The possibility that $m_0 = 0$ at the GUT
scale has recently been studied~\cite{eno5,emy}, and shown to be excluded by 
the phenomenological constraints. However, it was recalled that the input
supergravity scale could be somewhat higher than the GUT scale, in which
case one might find $m_0 \ne 0$ already at the GUT scale. Clearly the same
could also be true for $A_0$ and $B_0$. However, the deviations from
(\ref{noscalesusyx})  are model-dependent, and we think it important to
be aware of the phenomenological fate of the clear-cut $A_0 = 
B_0 = 0$ option for supersymmetry breaking.

\section{Electroweak Vacuum Conditions}

Before discussing the phenomenological constraints on this model, we first
show more precisely how the relation between $A$ and $B$ can be used to
determine $\tan \beta$ when the radiative electroweak symmetry breaking
conditions are applied.

In general, we start with the following set of input parameters defined at
the GUT scale: $m_{1/2}$, $m_0$, $A_0$, $B_0$ and the Higgs mixing 
parameter $\mu_0$.  By running the
full renormalization-group equations (RGEs) down to the weak scale and
minimizing the Higgs potential, one can solve for the Higgs vevs and
masses or, equivalently, $M_Z$, $\tan \beta$, and $m_A$. At the tree
level, these solutions take the simple form:
\ba
M_Z^2 & = & {2 (m_1^2 + \mu^2 - (m_2^2 + \mu^2) \tan^2 \beta) \over
(\tan^2 \beta -1)} \nonumber \\ 
\sin 2 \beta & = & { - 2 B \mu}/(m_1^2 + m_2^2 + 2 \mu^2) \nonumber \\ 
m_A^2 & = & m_1^2 + m_2^2 + 2 \mu^2 
\label{treerel}
\ea
where $m_1$ and $m_2$ are the soft supersymmetry-breaking masses for the
two Higgs doublets at the electroweak scale. However, since $M_Z$ is
known, and because the full one-loop set of tadpole equations does not
admit an analytical solution for $\tan \beta$, it is customary to use
$M_Z$ and $\tan \beta$ as inputs and instead solve for $\mu$ and $B$:
\ba
\mu^2 & = & \frac{m_1^2 - m_2^2 \tan^2 \beta + \frac{1}{2} \mz^2 (1 -
\tan^2 \beta) + \Delta_\mu^{(1)}}{\tan^2 \beta - 1 + \Delta_\mu^{(2)}} 
\nonumber \\
B \mu  & = & -{1 \over 2} (m_1^2  + m_2^2 + 2 \mu^2) \sin 2 \beta + \Delta_B
\label{onelooprel}
\ea
where $\Delta_B$ and $\Delta_\mu^{(1,2)}$ are loop
corrections~\cite{Barger:1993gh,deBoer:1994he,Carena:2001fw}, and here
$m_{1,2} \equiv m_{1,2}(\mz)$.  Since $\Delta_\mu$ depends on $\tan \beta$
and $\Delta_B$ depends on both $\mu$ and $\tan \beta$ in a nonlinear way,
it is not possible to write down an analytical solution for $\tan \beta$.  
The above set of inputs and outputs defines the CMSSM.

In the types of models discussed in the previous section, we have specific
GUT-scale
boundary conditions on $B_0$, namely $B_0 = A_0 - m_0$ in minimal
supergravity models or $B_0 = A_0 = 0$ in no-scale models. Therefore, we
cannot treat the value of $B(M_Z)$ as a free parameter, and instead must
solve numerically for $\tan \beta$. Thus, a given value of $m_{1/2}$,
$m_0$, $A_0/m_0$, and $sgn(\mu)$ will correspond to a definite value for
$\tan \beta$.  When combined with the phenomenological constraints
discussed below, we can determine for a particular model of supersymmetry
breaking the allowed (and often quite restricted) values of $\tan \beta$.

\section{Phenomenological Constraints on $m_{1/2}$ and $m_0$}

We apply the standard LEP constraints on the supersymmetric parameter
space, namely $m_{\chi^\pm} > 104$~GeV \cite{LEPsusy}, $m_{\tilde e} > 99$~GeV 
 \cite{LEPSUSYWG_0101} and $m_h >
114$~GeV \cite{LEPHiggs}. The former two constrain $m_{1/2}$ and $m_0$ directly
via the sparticle masses, and the latter indirectly via the sensitivity of
radiative corrections to the Higgs mass to the sparticle masses,
principally $m_{\tilde t, \tilde b}$~\footnote{We assume as our default
that $m_t = 175$~GeV.}. We use the  latest version
of {\tt FeynHiggs}~\cite{FeynHiggs} for the calculation of $m_h$. We require
the branching ratio for $b \rightarrow
s \gamma$ to be consistent with the experimental measurements \cite{bsg}. We also
indicate the regions of the $(m_{1/2}, m_0)$ plane that are favoured by
the BNL measurement ~\cite{newBNL} of $g_\mu - 2$ at the 2-$\sigma$ level, corresponding to 
a deviation of $(33.9 \pm 11.2) \times 10^{-10}$ from the Standard Model calculation
of~\cite{Davier} using $e^+ e^-$ data.  We are however 
aware that this constraint is still under discussion and do not use it to
constrain $\tan \beta$. All the $\mu > 0$
planes would be consistent with $g_\mu - 2$ at the 3-$\sigma$ level,
whereas $\mu < 0$ is disfavoured even if one takes a relaxed view of the
$g_\mu - 2$ constraint.

Finally, we impose the following requirement on the relic density of
neutralinos $\chi$: $0.094 \le \Omega_\chi h^2 \le 0.129$, as suggested by
the recent WMAP data~\cite{wmap}, in agreement with earlier indications.
We recall that several cosmologically-allowed domains of the $(m_{1/2},
m_0)$ planes for different values of $\tan \beta$ have been discussed
previously in the general CMSSM framework 
\cite{efgos} - \cite{efgosi}, \cite{otherOmega} - \cite{hyperbolic}. One is a
`bulk' region at low
$m_{1/2}$ and $m_0$, which has been squeezed considerably by the WMAP
constraint on $\Omega_\chi h^2$. A second region is the $\chi - {\tilde
\tau_1}$ coannihilation `tail' \cite{stauco,moreco}, which stretches to larger
$m_{1/2}$, close
to the boundary of the acceptable region where $m_\chi \le m_{\tilde
\tau_1}$. In the wake of WMAP, this `tail' is now much narrower - because
of the smaller range of $\Omega_\chi h^2$ - and shorter - because of the
more stringent upper limit on $\Omega_\chi h^2$ \cite{eoss,wmapothers}. A third
region is the
`funnel' due to rapid $\chi \chi \to H, A$ annihilation that occurs at
larger $m_0$ and $m_{1/2}$ \cite{efgosi,funnel}. Finally, the fourth domain is the
`focus-point' region at large $m_0$, close to the boundary where radiative
breaking of electroweak symmetry is no longer possible \cite{focus,hyperbolic}.

We see in the next Section that the `funnel' and `focus-point' regions are
not present in the simple models of supersymmetry breaking
introduced earlier, whilst the `bulk' region is possible only for a very
restricted range of $\tan \beta$. On the other hand, the coannihilation
`tail' generally remains permitted.

\section{Examples of $(m_{1/2}, m_0)$ Planes}

\begin{figure}
\begin{center}
\mbox{\epsfig{file=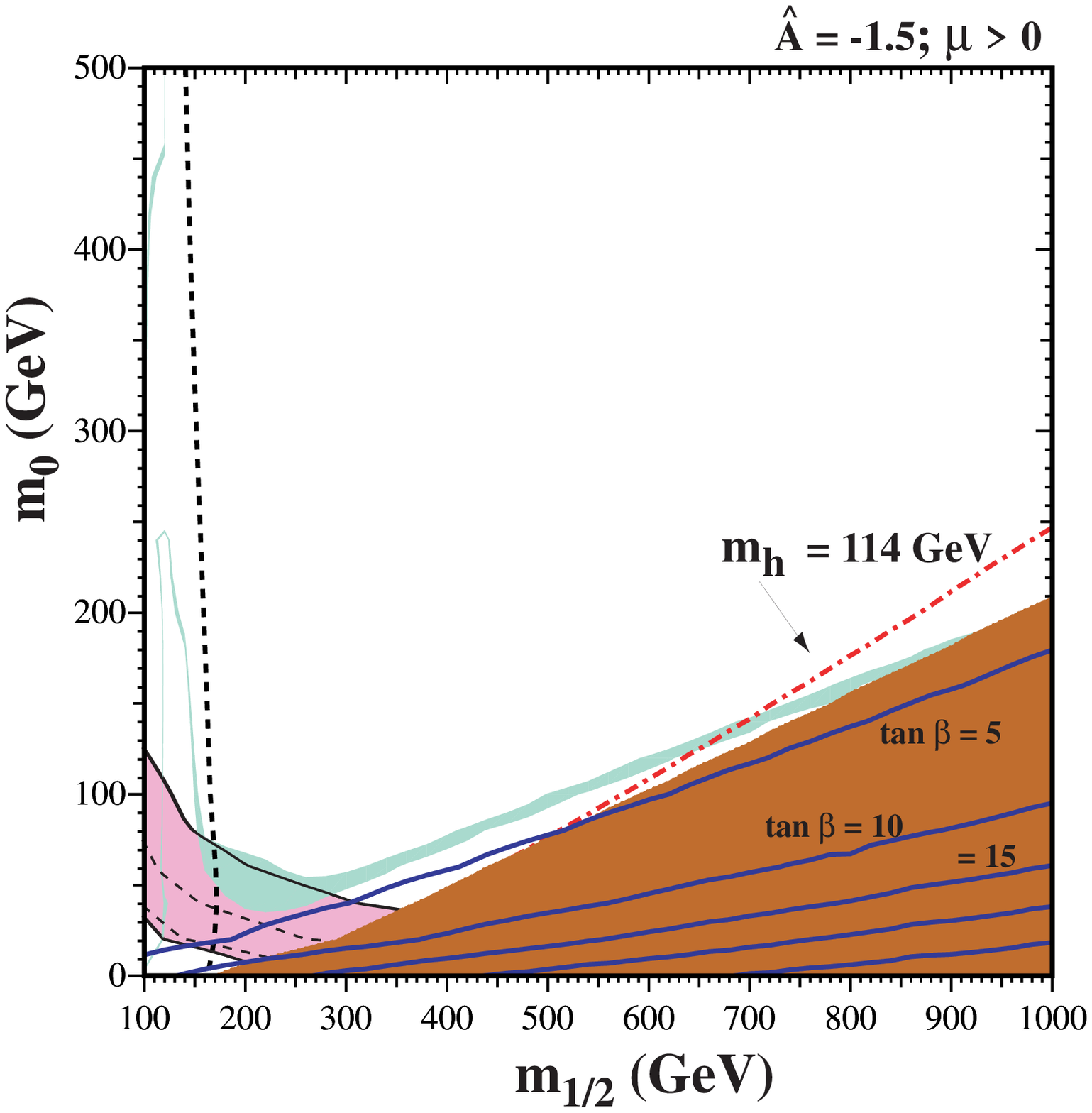,height=7cm}}
\mbox{\epsfig{file=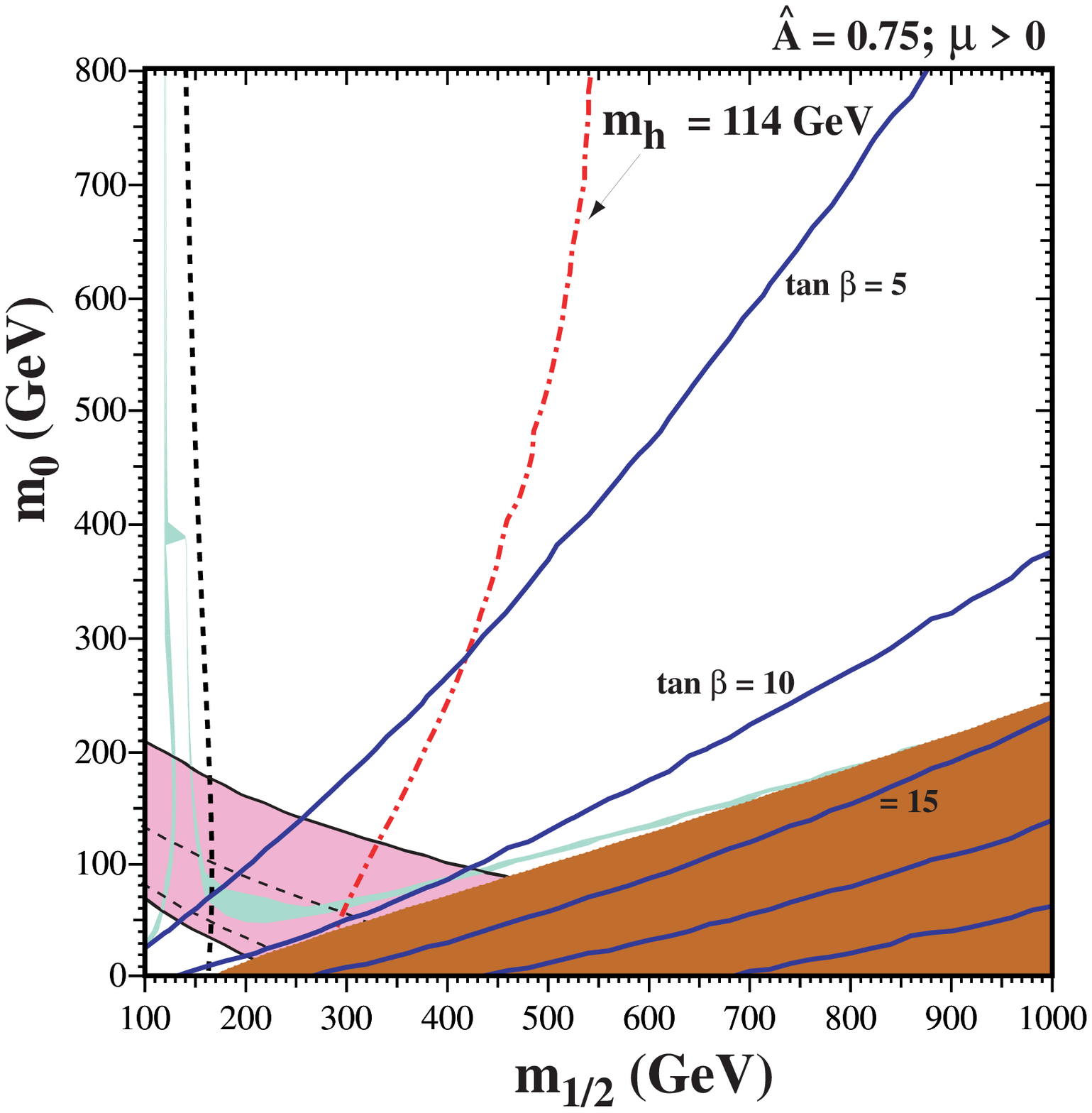,height=7cm}}
\end{center}
\begin{center}
\mbox{\epsfig{file=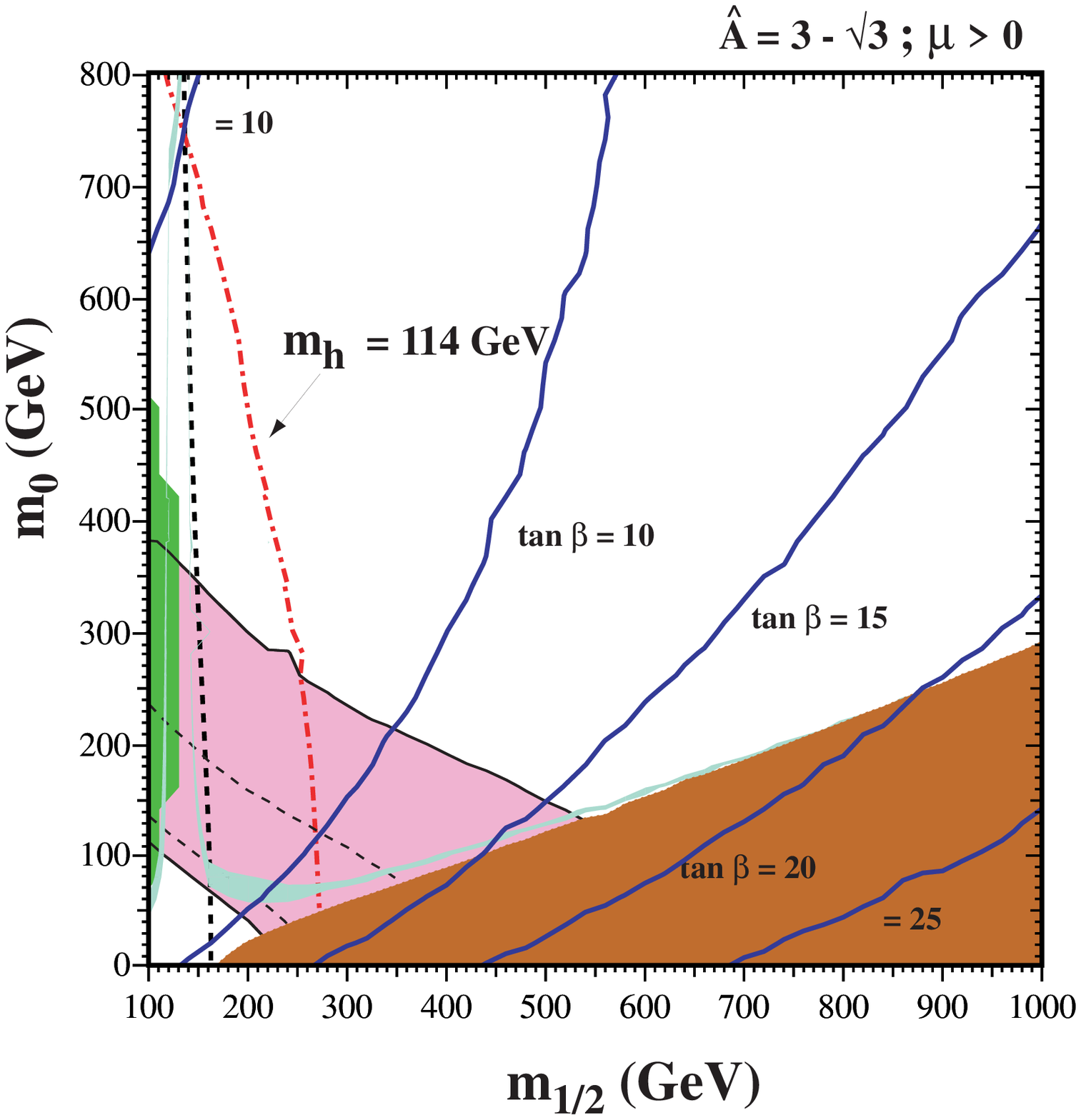,height=7cm}}
\mbox{\epsfig{file=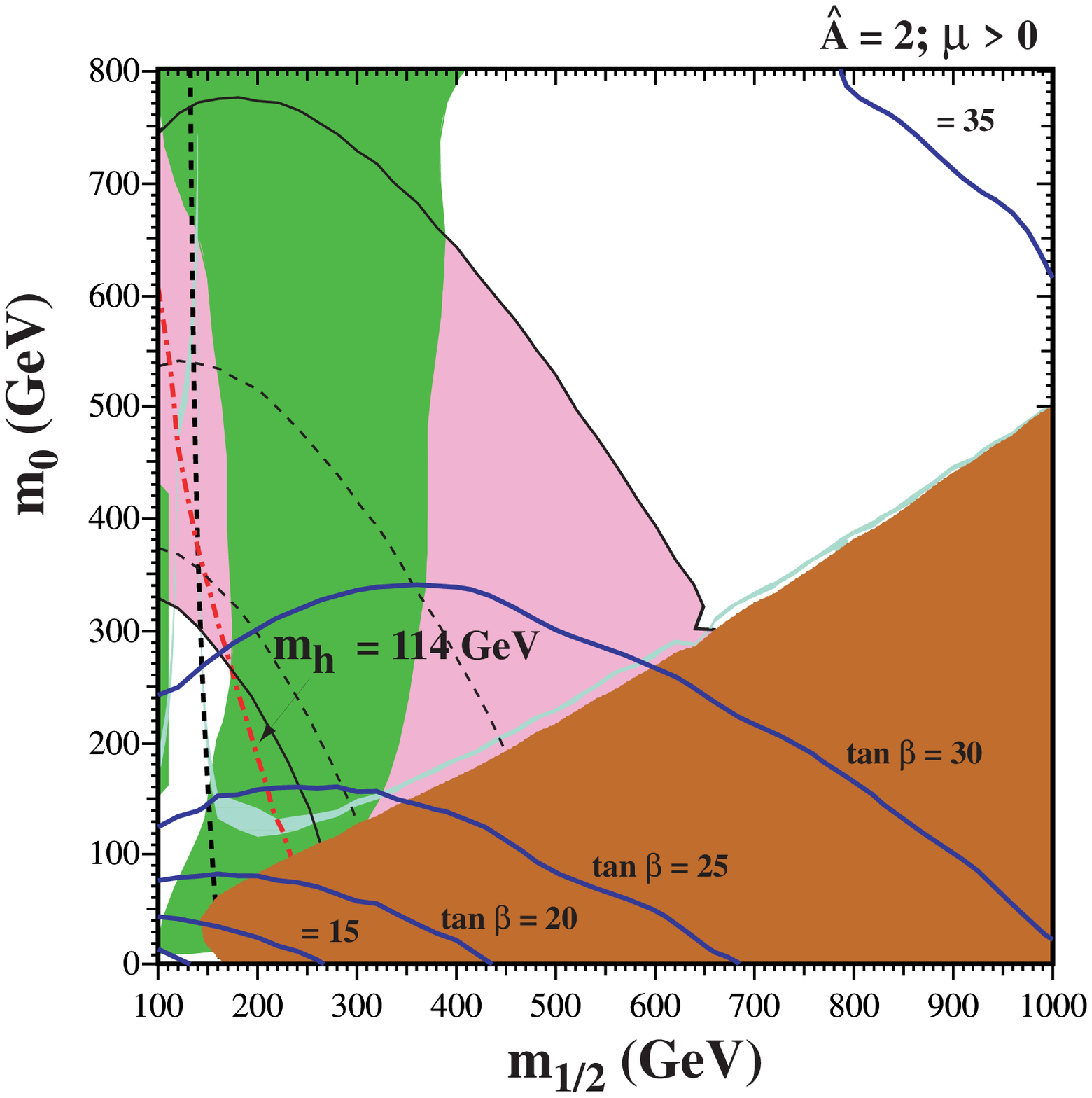,height=7cm}}
\end{center}
\caption{\it
Examples of $(m_{1/2}, m_0)$ planes with contours of $\tan \beta$ 
superposed, for $\mu > 0$ and (a) ${\hat A} = - 1.5, {\hat B} = 
{\hat A} -1$, (b) ${\hat A} = 0.75, {\hat B} =
{\hat A} -1$, (c) the simplest Polonyi model with ${\hat A} = 3 - 
\sqrt{3}, {\hat B} = {\hat A} -1$ and (d) ${\hat A} = 2.0, {\hat B} =
{\hat A} -1$. In each panel, we show the regions excluded by 
the LEP lower limits on MSSM particles, those ruled out by $b
\to s \gamma$ decay~\protect\cite{bsg} (medium green shading), and those 
excluded 
because the LSP would be charged (dark red shading). The region favoured 
by the WMAP range $\Omega_{CDM} h^2 =
0.1126^{+0.0081}_{-0.0091}$ has light turquoise shading. The region 
suggested by $g_\mu - 2$ is medium (pink) shaded.}
\label{fig:Polonyi}
\end{figure}

We display in Fig.~\ref{fig:Polonyi} the contours of $\tan \beta$ (solid
blue lines) in the $(m_{1/2}, m_0)$ planes for selected values of ${\hat
A}$, ${\hat B}$ and the sign of $\mu$. Also shown are the contours where
$m_{\chi^\pm} > 104$~GeV (near-vertical black dashed lines) and $m_h >
114$~GeV (diagonal red dash-dotted lines). The excluded regions where
$m_\chi > m_{\tilde \tau_1}$ have dark (red) shading, those excluded by $b
\to s \gamma$ have medium (green) shading, and those where the relic
density of neutralinos lies within the WMAP range $0.094 \le \Omega_\chi
h^2 \le 0.129$ have light (turquoise) shading. Finally, the regions
favoured by $g_\mu - 2$ at the 2-$\sigma$ level are medium (pink) shaded.

As seen in panel (a) of Fig.~\ref{fig:Polonyi}, when $\mu > 0$ and ${\hat
A} = -1.5$, close to its minimum possible value, the contours of $\tan
\beta$ rise diagonally from low values of $(m_{1/2}, m_0)$ to higher
values, with higher values of $\tan \beta$ having lower values of $m_0$
for a given value of $m_{1/2}$. The $m_h = 114$~GeV contour rises in a
similar way, and regions above and to the left of this contour have 
$m_h < 114$ GeV and are excluded.  Therefore,  only a very limited 
range of $\tan \beta \sim 4$ is
compatible with the $m_h$ and $\Omega_{CDM} h^2$ constraints. 
At lower values of ${\hat A}$, the slope of the Higgs contour softens and 
even less of the parameter space is allowed.  Below ${\hat A} \simeq -1.9$, 
the entire $m_{1/2} - m_0$ plane is excluded. When ${\hat
A}$ is increased to 0.75, as seen in panel (b) of Fig.~\ref{fig:Polonyi},
both the $\tan \beta$ and $m_h$ contours rise more rapidly with $m_{1/2}$,
and a larger range $ 9 \la  \tan \beta \la 14$ is allowed~\footnote{Note that the
contours for given values of $\tan \beta$ always intersect the axis $m_0 =
0$ at the same value of $m_{1/2}$.}. In the simplest Polonyi model with
${\hat A} = 3 - \sqrt{3}$ shown in panel (c) of Fig.~\ref{fig:Polonyi}, we
see that the $\tan \beta$ contours have noticeable curvature. In this
case, the Higgs constraint combined with the relic density requires $\tan \beta
\gappeq 11$, whilst the relic density also enforces $\tan \beta \lappeq 
20$~\footnote{The other Polonyi case with ${\hat A} = - 3 + \sqrt{3}$ 
(not shown) is very similar to panel (a) for ${\hat A} = - 1.5$, and has 
a very narrow allowed range of $\tan \beta \sim 4.5$.}. Finally, in panel 
(d) of
Fig.~\ref{fig:Polonyi}, when ${\hat A} = 2.0$, close to its maximal value
for $\mu > 0$, the $\tan \beta$ contours turn over towards smaller
$m_{1/2}$, and only relatively large values $25 \la \tan \beta \la 35$ are
allowed by the $b \to s \gamma$ and $\Omega_{CDM} h^2$ constraints,
respectively.

\begin{figure}
\begin{center}
\mbox{\epsfig{file=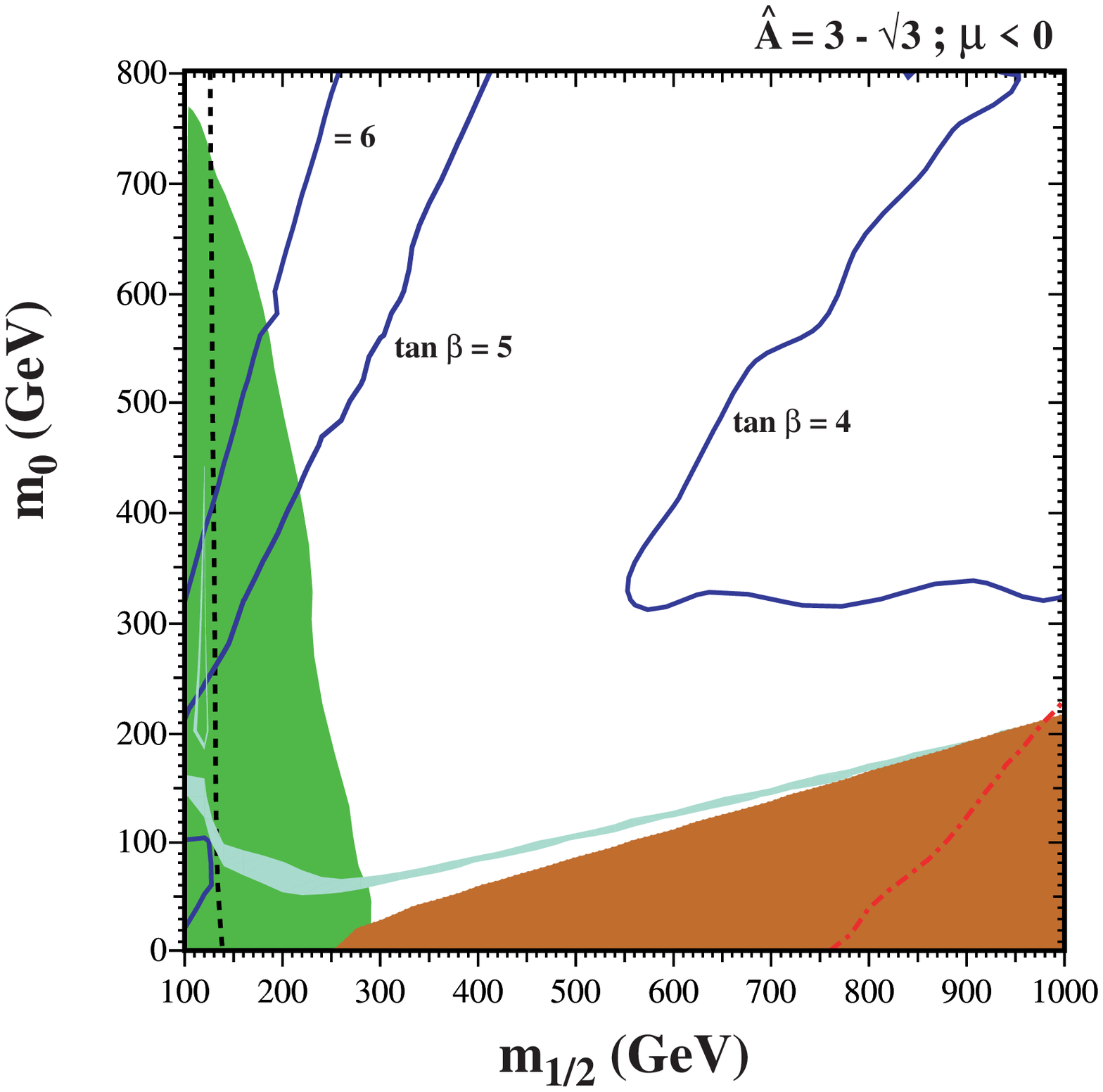,height=7cm}}
\mbox{\epsfig{file=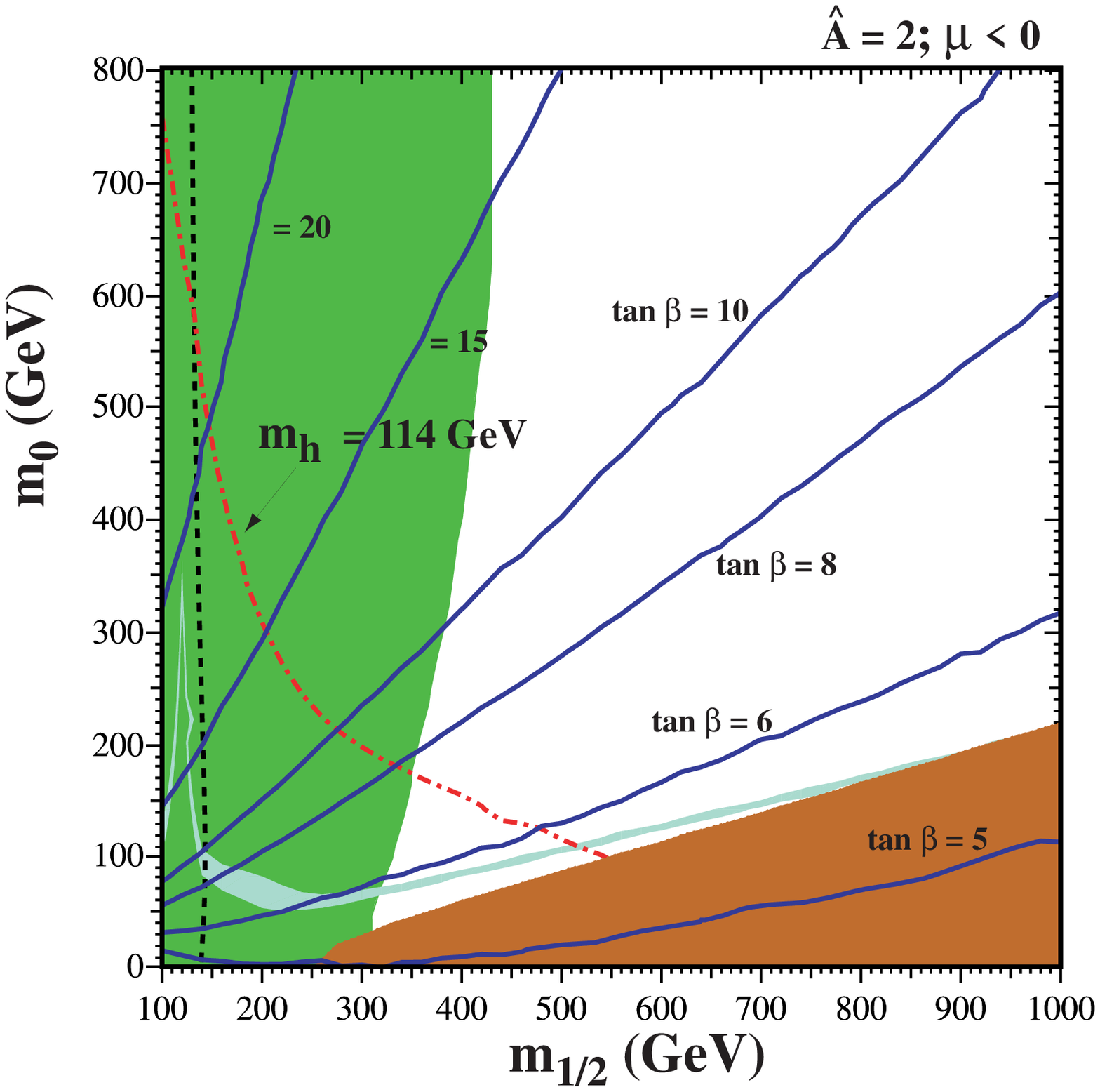,height=7cm}}
\end{center}
\caption{\it
As in Fig.~\ref{fig:Polonyi}, but now for $\mu < 0$ and the 
choices (a) ${\hat A} = 3 - \sqrt{3}, {\hat B} = 
{\hat A} -1$ and (b) ${\hat A} = 2, {\hat B} =
{\hat A} -1$ and $\mu < 0$. 
}
\label{fig:Polonyin}
\end{figure}

In the case of $\mu < 0$, negative values of ${\hat A}$ are not allowed,
and only a tiny area in the $(m_{1/2}, m_0)$ plane near the end point of
the coannihilation tail around $m_{1/2} = 1000$ GeV is allowed in the
positive Polonyi case ${\hat A} = 3 - \sqrt{3}$, as seen in panel (a) of
Fig.~\ref{fig:Polonyin}. This is because the Higgs and $\Omega_{CDM} h^2$
constraints are barely compatible in this case, and allow only $\tan \beta
\simeq 4.15$. At larger values of ${\hat A}$, the allowed region is
extended, as exemplified in panel (b) of Fig.~\ref{fig:Polonyin} for the
case ${\hat A} = 2.0$, where a small region around $\tan \beta \simeq 5.5
- 5.7$ is allowed. This panel shows that, approximately, the value of
$\tan \beta$ depends only on the ratio $m_{1/2}/m_0$~\cite{th}.

There are several generic patterns in the results above that can be
explained qualitatively, as follows. First, we notice that for any given
value of $(m_{1/2}, m_0)$, $\tan \beta$ increases as ${\hat A}$ increases.  
The reason for this can be found by looking at the second equation of
(\ref{treerel}), and setting $A_0 = B_0 + m_0$. For large $\tan \beta$,
$\sin 2 \beta \sim 1/\tan \beta$, so $B$ at the weak scale is inversely
proportional to $\tan \beta$, at the tree level. In the $\mu > 0$ case,
this tree-level value of $B$ is negative, so its value {\it grows} as
$\tan \beta$ increases. While loop corrections are generally negative 
for $\mu > 0$, and RGE corrections to obtain $B(M_X)$ are positive, the
monotonic growth of $B_0$ with $\tan \beta$ is preserved. Thus the
resulting value of $B_0$, and hence also $A_0$, increases with $\tan
\beta$. In the $\mu < 0$ case, the tree-level value of $B$ is generally positive
(the exception being when $m_1^2 + m_2^2 + 2 \mu^2 < 0$), and so its value
{\it decreases} as $\tan \beta$ increases. However, there are some terms
in the loop correction $\Delta_B$ that are proportional to $\mu \tan
\beta$ and flip the sign of $\Delta_B$ at a particular value of $\tan
\beta$, so that the full one-loop $B(M_W)$ is then again an increasing
function of $\tan \beta$, and likewise $A_0$.

Using similar arguments, we can further understand the different
behaviours of the $\tan\beta$ contours when $\mu$ is positive or negative
with fixed ${\hat A}$, for example in the last panels in
Fig.~\ref{fig:Polonyi} and Fig.~\ref{fig:Polonyin} for ${\hat A}=2$. To
this end, look at the second equation in (\ref{onelooprel}), bearing in
mind that $\sin 2 \beta \sim 1/\tan \beta$. For $\mu>0$ and fixed $m_0$,
as $m_{1/2}$ increases both $\Delta_B$ and the RGE corrections to $B$
increase, yielding a relatively constant value for $\tan\beta$ when
the growth of the term $-\Delta_B$ almost compensates the positive
RGE corrections. For large values of $m_{1/2}$, the RGE corrections take
over, resulting in the bending of the $\tan\beta$ contours. On the other
hand, for $\mu<0$, the flipping of the sign of $\Delta_B$ described in the
paragraph above results in different behaviour. In this case, as $m_{1/2}$
increases with fixed $m_0$, $\tan\beta$ always decreases.

In panel (a) of  Fig.~\ref{fig:Polonyin}, the magnitude of the tree 
level value of $B$ at
the weak scale increases with $m_0$, decreasing the value of $\tan
\beta$. However, the loop correction is also growing, tending to
increase  $\tan \beta$. 
We see from the figure that $\tan \beta$ is first decreasing and
then increasing  as $m_0$ is increased. This behaviour is different 
from panel
(b) of  Fig.~\ref{fig:Polonyin}, where the tree level value of $B$ at
the weak scale is decreasing with $m_0$, and dominates the 
determination of
$\tan \beta$, which is now increasing monotonically. 

At high values of ${\hat A}$ (and high $\tan \beta)$, the off-diagonal 
elements  in the squark mass matrix become large at large $m_0$.
Therefore, we find no solutions which are phenomenologically viable
above a certain value of ${\hat A}$. This is
because the regions where the LSP is the $\stau$ or the $\stop$ close off
the parameter space~\footnote{The neutralino-stop coannihilation region
which occurs when $A_0$ is large in the small $(m_{1/2}, m_0)$
region~\cite{stopco} does not appear in our analysis because $A_0$ is
still too small.}. In fact, this feature is generic in the CMSSM as shown in
Fig. 3 of \cite{efgo}. This effect is more severe at large $\tan\beta$, which 
further compounds the difficulty in going to large values of ${\hat A}$ in the type of models
discussed here.

Finally, we note the absences of both the funnel and the focus-point
regions. In the case of the funnel, this is due to the relatively small values of
$\tan \beta$ allowed in the class of models considered here: we recall
that the funnel region appears only for large $\tan \beta \gappeq 45$ for
$\mu > 0$ and $\tan \beta \gappeq 30$ for $\mu < 0$ in the CMSSM.

To understand the absence of the focus-point region, we refer
to~\cite{hyperbolic}, where it was shown that the position of the focus
point is sensitive to the value of $A_0$. As $A_0$ is increased, the focus
point is pushed up to higher values of $m_0$. Here, with $A_0 \propto m_0$, the
focus-point region recedes faster than $m_0$ if ${\hat A}$ is large
enough, and is therefore never encountered. For small ${\hat A}$, $\tan
\beta$ is small at large $m_0$, as shown in panel (b) of
Fig.~\ref{fig:Polonyi}, so we do not find a focus point in this case,
either. In addition, as can be inferred from the small disconnected
segment of the $\tan \beta = 10$ contour in the top left corner of panel
(c), all the $\tan \beta$ contours loop back down to lower $m_0$ before
reaching the focus-point region.

The above analysis shows that the `bulk' $\Omega_{CDM} h^2$ region is
almost completely excluded by the Higgs constraint, but a larger fraction
would be allowed if we allowed a 2-GeV error in the CMSSM Higgs mass
calculation, or if $m_t$ turns out to be significantly greater than
$175$~GeV. Almost all the coannihilation `tail' region is allowed. As remarked on above, there
is no `funnel' region at large $m_{1/2}$ and $m_0$, nor any `focus-point'
region at large $m_0$.

\section{Bounds on $\tan \beta$}

\begin{figure}
\begin{center}
\mbox{\epsfig{file=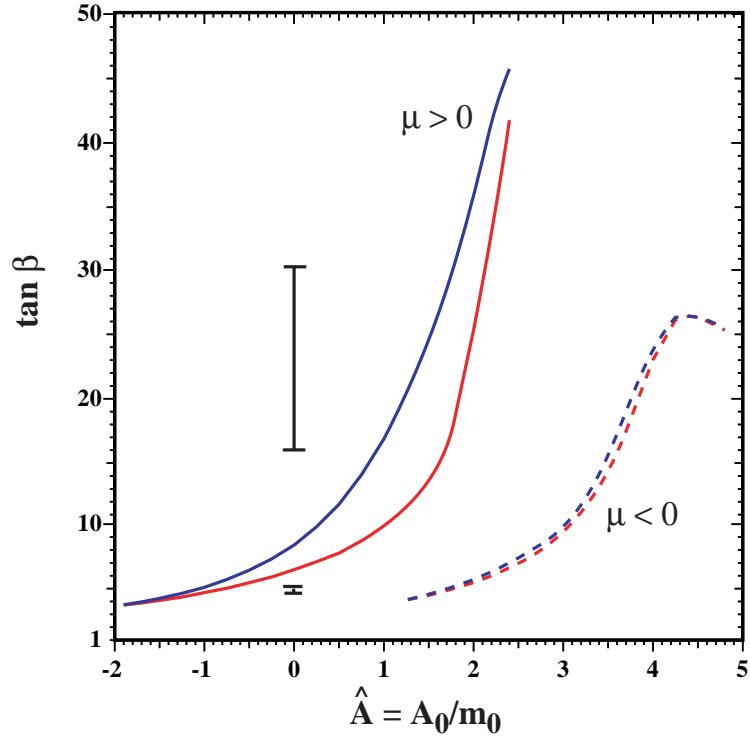,height=10cm}}
\end{center}
\caption{\it
The ranges of $\tan \beta$ allowed if ${\hat B} = {\hat A} - 1$ for $\mu > 
0$ (solid lines) and $\mu < 0$ (dashed lines). The Polonyi model
corresponds to ${\hat A}  \simeq \pm 1.3$. Also shown as `error bars' 
are the ranges of $\tan \beta$ allowed in the no-scale case ${\hat A} = 
{\hat B} = 0$ for $\mu > 0$ (upper) and $\mu < 0$ (lower).}
\label{fig:tanbeta}
\end{figure}

It is clear from the previous figures that only limited ranges of $\tan 
\beta$ are consistent with the phenomenological constraints within any 
given pattern of supersymmetry breaking. We display in 
Fig.~\ref{fig:tanbeta} the ranges of $\tan \beta$ allowed as a function  
of ${\hat A}$. For ${\hat B} = {\hat A} -1$ and $\mu > 0$, as shown by the 
solid lines, we see that the upper and lower limits on $\tan \beta$ both 
increase monotonically with ${\hat A}$. We find consistent solutions to 
all the phenomenological constraints only for
\beq
- 1.9 \; < \; {\hat A} \; < \; 2.5,
\label{rangeA}
\eeq
over which range
\beq
3.7 \; < \; \tan \beta \; \la \; 46.
\label{rangetb}
\eeq
Generally speaking, the range of $\tan \beta$ for any fixed value of 
${\hat A} < 0$ is very restricted, with larger ranges of $\tan \beta$ 
becoming allowed for ${\hat A} > 0$. In the specific case of the simplest 
Polonyi model with positive ${\hat A} = 3 - \sqrt{3}$, we find
\beq
11 \; < \; \tan \beta \; < \; 20,
\label{Polonyitb}
\eeq
whereas the range in $\tan \beta$ for the negative Polonyi model 
with ${\hat A} = \sqrt{3} - 3$, is 4.4 -- 4.6.
Furthermore, the difference between the upper and lower limits on  $\tan \beta$ 
never exceeds $\sim$ 14 for any fixed value of  ${\hat A}$.

The corresponding results for $\mu < 0$ are 
\beq
1.2 \; < \; {\hat A} \; < \; 4.8,
\label{rangeAn}
\eeq
over which range
\beq
4 \; < \; \tan \beta \; \la \; 26.
\label{rangetbn}
\eeq
The range of ${\hat A}$ is shifted, and the range of $\tan \beta$ reduced, 
as compared to the case of $\mu > 0$. In particular, the negative Polonyi 
model is disallowed and the positive version is allowed only for $\tan 
\beta \sim 4.15$.

\section{No-Scale Models}

\begin{figure}
\begin{center}
\mbox{\epsfig{file=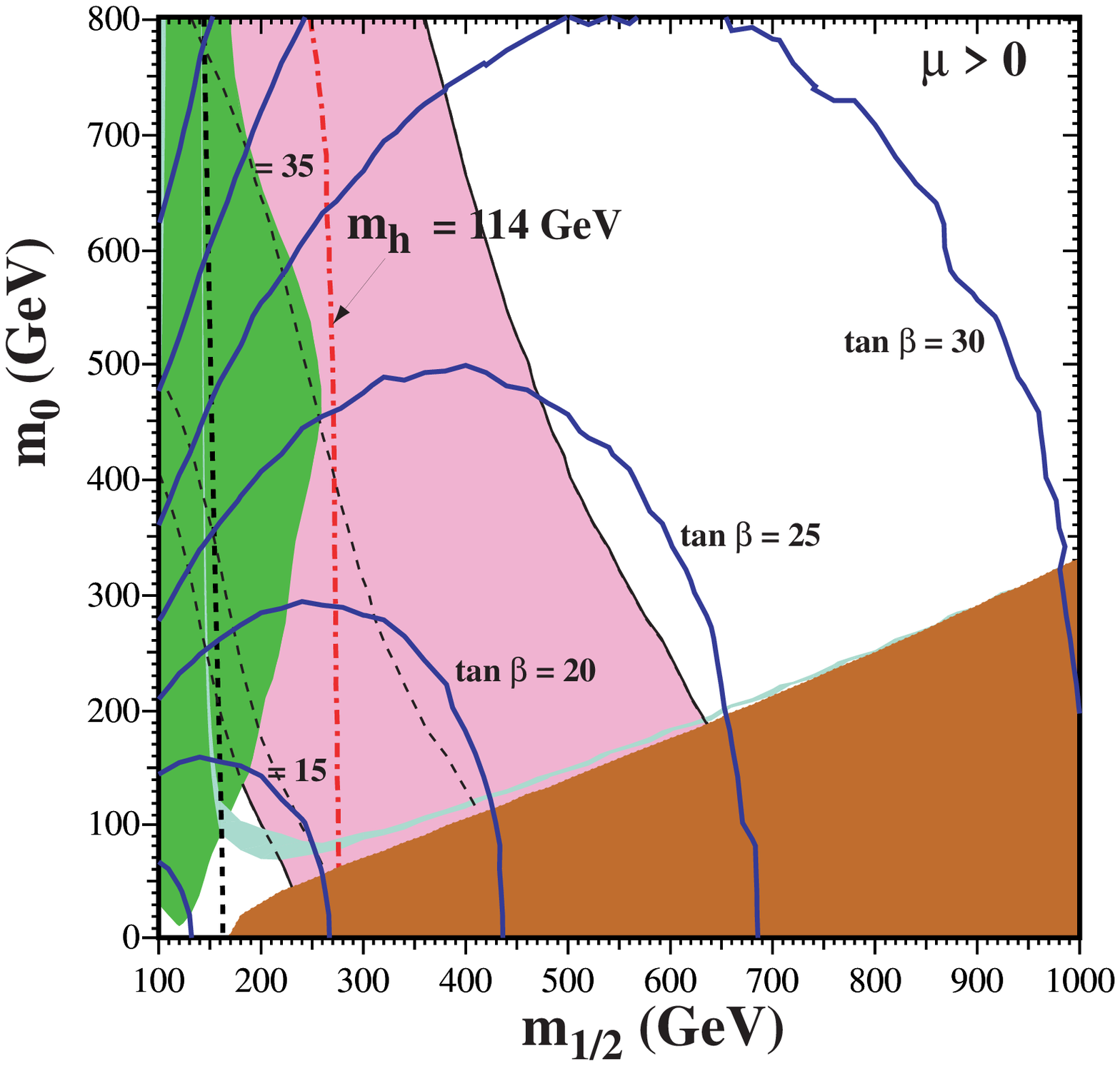,height=7cm}}
\mbox{\epsfig{file=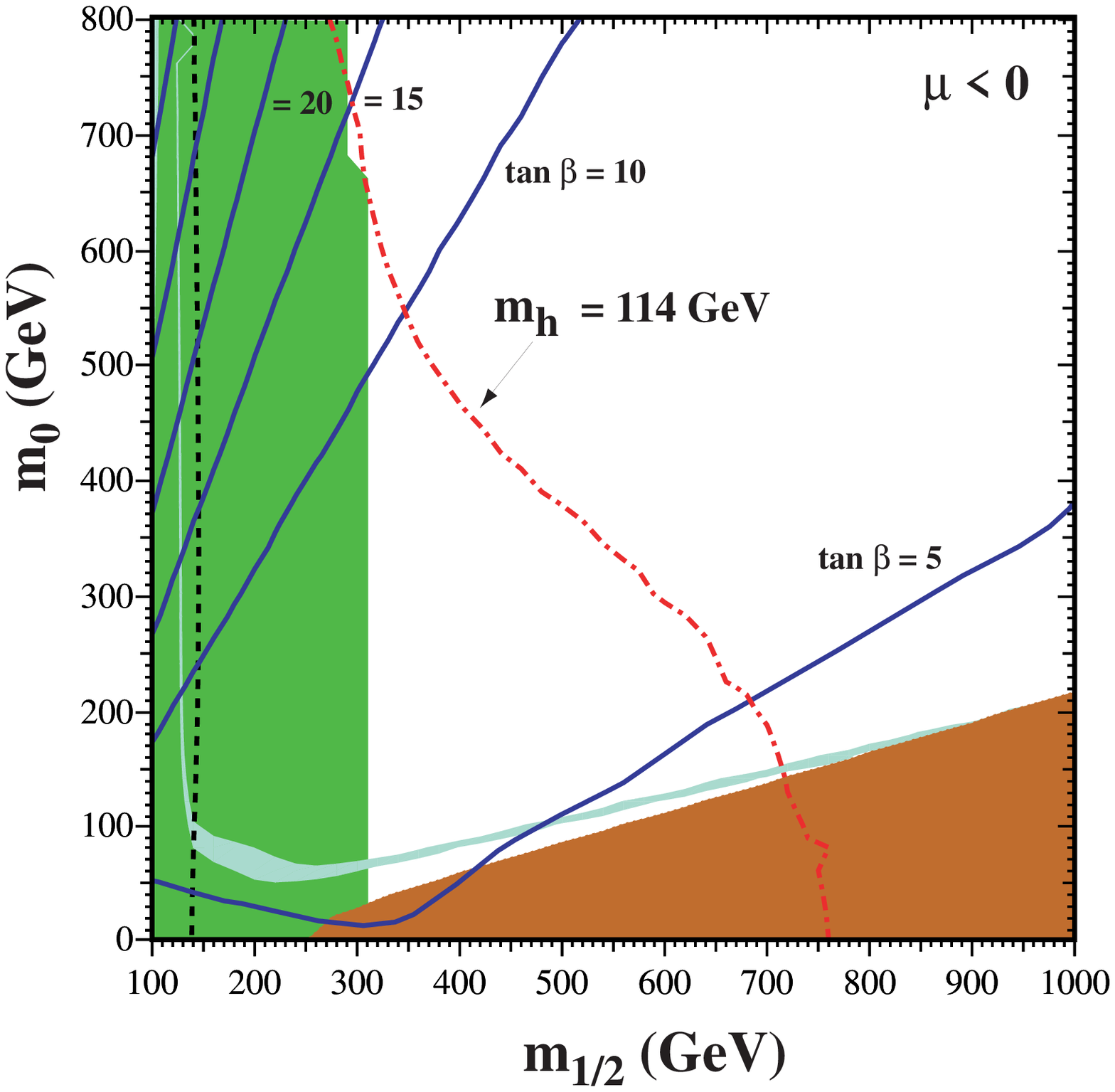,height=7cm}}
\end{center}
\caption{\it
As in Fig.~\ref{fig:Polonyi}, for the no-scale cases ${\hat A} = 0,
{\hat B} = 0$ and (a) $\mu > 0$, (b) $\mu < 0$.
}
\label{fig:noscale}
\end{figure}

We display in Fig.~\ref{fig:noscale} the results of a similar analysis for 
the no-scale case ${\hat A} = {\hat B} = 0$. For $\mu > 0$, the allowed 
range of $\tan \beta$ is
\beq
16 \; < \; \tan \beta \; < 30, 
\label{tbpnoscale}
\eeq
where the lower limit is provided by the Higgs search, and the upper limit
is at the tip of the coannihilation `tail'. For $\mu < 0$, the same
constraints allow just a small range around $\tan \beta \sim 4.8$.  These
two ranges are both shown as `error bars' in Fig.~\ref{fig:tanbeta}.

However, the other no-scale condition $m_0 = 0$ is not allowed for either
sign of $\mu$, the minimum being $m_0 \simeq 62$~GeV for $\mu > 0$ and
$\tan \beta \simeq 16$. The fact that $m_0 \ne 0$ is no surprise, since the 
same conclusion was reached previously without imposing the supplementary
no-scale conditions ${\hat A} = {\hat B} = 0$~\cite{eno5}. However, as we
have already pointed out, the no-scale boundary conditions should be
interpreted as applying at the supergravity scale, so it is possible that
$m_0, {\hat A}, {\hat B}$ all $\ne 0$, albeit small, at the GUT scale.
We note that in this case, there is in fact a focus-point region at 
roughly the same position as in the CMSSM with $A_0 = 0$.

\section{Conclusions}

We have shown in this paper that only a restricted range of $\tan \beta$
is allowed in any specific pattern of supersymmetry breaking. We have
illustrated this point by discussions of minimal supergravity models with
${\hat A} = {\hat B} + 1$ and no-scale models with ${\hat A} = {\hat B} =
0$, but the same comment would apply to other models of supersymmetry
breaking not discussed here. Within the class of minimal supergravity
models, we have selected in particular the simplest Polonyi model with
$\vert {\hat A} \vert = 3 - \sqrt{3}$, but also discussed models with
other values of ${\hat A}$, finding a rather restricted range, in
particular for $\mu < 0$.

One inference from our analysis is that an experimental determination of 
$\tan \beta$ could be a useful discriminator between different models of 
supersymmetry breaking. To understand the potential scope of this analysis 
tool, it would be necessary to study a wider class of models of 
supersymmetry breaking than those discussed here.

\vskip 0.5in
\vbox{
\noindent{ {\bf Acknowledgments} } \\
\noindent  The work of K.A.O., Y.S., and V.C.S. was supported in part
by DOE grant DE--FG02--94ER--40823.}

\end{document}